\newif\ifCLNS \CLNStrue

\ifCLNS  
 \documentstyle[preprint,aps]{revtex} 
\else
 \documentstyle[prl,aps]{revtex}      
\fi      

\ifCLNS  
\fi      

\begin{document}

\draft 

\ifCLNS  
\tighten 
\preprint{\vbox{\hbox{CLNS 94/1281 \hfill}
                \hbox{CLEO 94--11  \hfill}
                }}
\fi      


\title{A Measurement of the Branching Fraction
${\cal B}(\tau^\pm\rightarrow h^\pm\pi^0\nu_\tau)$ }

\author{
M.~Artuso,$^{1}$ M.~Goldberg,$^{1}$ D.~He,$^{1}$ N.~Horwitz,$^{1}$
R.~Kennett,$^{1}$ R.~Mountain,$^{1}$ G.C.~Moneti,$^{1}$
F.~Muheim,$^{1}$ Y.~Mukhin,$^{1}$ S.~Playfer,$^{1}$
% ...
}

\address{
\ifCLNS  
\bigskip 
\fi      
{\rm (CLEO Collaboration)}\\  
\ifCLNS  
\newpage 
\fi      
$^{1}${Syracuse University, Syracuse, New York 13244}\\
% ...
\ifCLNS  
\bigskip 
\fi      
}        


\maketitle

\begin{abstract}
Using data from the CLEO II detector at CESR,
we measure ${\cal B}(\tau^\pm\rightarrow h^\pm\pi^0\nu_\tau)$
where $h^\pm$ refers to either $\pi^\pm$ or $K^\pm$.
We use three different methods to measure this branching fraction.
The combined result is
${\cal B}(\tau^\pm\rightarrow h^\pm\pi^0\nu_\tau)
= 0.2587 \pm 0.0012 \pm 0.0042$,
in good agreement with Standard Model predictions.
This result, in combination with other precision measurements,
reduces the significance of the one-prong problem in tau decays.
\end{abstract}



\narrowtext 


The largest decay mode of the tau lepton
is to one charged pion, one $\pi^0$, and a neutrino.
The branching fraction for this mode is
not precisely known,
with measured values ranging from 22\%\ to 26\%\ \cite{ref:PDG,ref:recent}.
Both the magnitude and shape of the $\pi^\pm\pi^0$ spectrum
provide information on the charged weak spectral function
and a test of the Conserved Vector Current (CVC) hypothesis.
The significance of the ``one-prong problem'' \cite{ref:PDG,ref:onepr},
the gap between the inclusive one-prong branching fraction
and the sum of exclusive modes,
depends strongly on the magnitude and precision of the
$\tau^\pm\rightarrow h^\pm\pi^0\nu_\tau$ branching fraction
(referred to as the $h^\pm\pi^0\nu_\tau$ mode;
$h^\pm$ refers to either $\pi^\pm$ or $K^\pm$).
In addition, the CLEO
measurements of the multi-$\pi^0$ ($h^\pm n\pi^0\nu_\tau$, $n>1$)
decay branching fractions \cite{ref:npio}
are normalized to the
$h^\pm\pi^0\nu_\tau$ mode in order to minimize sytematic errors;
a precision measurement of the
$h^\pm\pi^0\nu_\tau$ mode is then needed to turn those
ratios into branching fractions.

In this paper we present a precise measurement of
the branching fraction
${\cal B}_{h\pi^0} \equiv {\cal B}(\tau^\pm\rightarrow h^\pm\pi^0\nu_\tau)$.
The $h^\pm\pi^0\nu_\tau$ mode is dominated by $\rho^\pm\nu_\tau$, but also
receives contributions from $K^*\nu_\tau$, $\rho^\prime\nu_\tau$,
and non-resonant $h^\pm\pi^0\nu_\tau$ modes.
This final state is referred to in this paper as the $\rho$ mode
for brevity.
No attempt is made to distinguish $\pi^\pm$ from $K^\pm$
in this analysis; the charged particles are assumed to be pions.

Taus are produced in pairs at $e^+e^-$ colliders,
and the tau which recoils against the one observed to decay
into $h^\pm\pi^0\nu_\tau$ is referred to as the tag.
We use four different tag decay modes:
$e^\pm \nu_e \nu_\tau$, 
$\mu^\pm \nu_\mu \nu_\tau$, 
$h^\pm\pi^0\nu_\tau$, 
and $h^\pm h^+ h^- (n\pi^0)\nu_\tau$, 
denoted as $e$, $\mu$, $\rho$, and $3$ tag, respectively.
We choose combinations of event topologies which yield
the branching fraction for $h^\pm\pi^0\nu_\tau$
with minimum systematic errors and uncorrelated statistical errors.

Denoting the measured number of background-subtracted, acceptance-corrected
events of a given topology by, {\it e.g.},
$N_{\mu\rho}$ for the $\mu$ {\it vs.}~$\rho$ final state,
we calculate the branching fraction via the following methods:
\begin{equation}
{\cal B}_{h\pi^0} = \sqrt{\frac{N_{e\rho} N_{\mu\rho}}{2 N_{e\mu}
                    N_{\tau\tau} }} , \quad
                \sqrt{\frac{N_{\rho\rho}}{N_{\tau\tau}}},
                 \quad \hbox{or} \quad
                \frac{N_{3-\rho}}{N_{3-1}} \; {\cal B}_1.
\end{equation}
These will be referred to as the
$\ell-\rho$, $\rho-\rho$, and $3-\rho$ methods, respectively.
The $\ell-\rho$ and $\rho-\rho$ methods normalize to the number of produced
tau pairs $N_{\tau\tau}$, determined from the
measured luminosity and the theoretical tau pair
production cross section, as discussed below;
the $3-\rho$ method is independent of these quantities.
Since the branching fraction for the first two methods is determined from the
square root of the measured rates, errors associated
with these rates and with $N_{\tau\tau}$ are halved.
The $\ell-\rho$ method arranges the measured rates
in a combination that is independent of both the
leptonic decay branching fractions of the tau
and the lepton identification efficiency.
The $\rho-\rho$ method has no dependence on branching fractions
other than the one that is being measured.
The $3-\rho$ method is designed to avoid the uncertainty
in the three-prong inclusive branching fraction
of the tau by normalizing instead to the one-prong inclusive
branching fraction ${\cal B}_1$.
The current world average for ${\cal B}_1$
is \cite{ref:PDG,ref:onepr} $0.852\pm0.004$.
All three methods depend on understanding the absolute
efficiency for reconstructing the $\pi^0$,
as well as the level of
background from other $\tau^+\tau^-$ topologies.


The data were accumulated using the CLEO II detector \cite{ref:CLEOd}
with the CESR $e^+e^-$ collider
at beam energy $E_b \sim 5.3$ GeV and
center-of-mass energy $E_{cm} = 2 E_b$.
The analysis uses information from a 67-layer tracking system and a
7800-crystal CsI calorimeter, both of which are
inside a 1.5 T superconducting solenoidal magnet,
and a muon identification system.


A total integrated luminosity of 1.58 fb$^{-1}$ is used for this analysis.
The luminosity is measured using $e^+e^-\rightarrow e^+e^-$,
$\mu^+\mu^-$, and $\gamma\gamma$ final states,
and is determined with an error of 1\% \cite{ref:lumin}.
The tau pair production cross section
$\sigma(\tau^+\tau^-)$
at these center-of-mass energies is calculated
(to order $\alpha^3$) to be 1.173 times the
order $\alpha^2$ cross section of $86.86/E_{cm}^2$ nb
($E_{cm}$ in GeV), with a theoretical
relative uncertainty of 1\%\ \cite{ref:korb}.
The total number of produced tau pairs is calculated
by summing the cross section times luminosity
for each CLEO data run; this yields
$N_{\tau\tau} = 1.440\times 10^{6}$ produced tau pairs, with
a systematic error of 1.4\%.


The event selection and $\pi^0$ reconstruction methods
are similar for all event topologies.
For the $e-\mu$, $e-\rho$, $\mu-\rho$, and $\rho-\rho$ topologies,
we require an event to contain exactly
two charged tracks
separated in angle by $>90^\circ$.
For the $3-\rho$ and $3-1$ topologies,
we require an event to contain exactly
four charged tracks
with three in one hemisphere (as defined by the event thrust axis).
The event must have zero net charge.

All charged tracks
are required to be observed
in the central part of the detector
($|\cos\theta|<0.71$ where $\theta$ is the angle of the track with respect
to the $e^+$ beam),
to be consistent with coming from the
nominal $e^+e^-$ collision point,
and to have momentum greater than $0.1 E_b$.
Calorimeter showers are considered as candidates
for photons from $\pi^0$ decays if they are observed
in the central part of the detector,
are not matched to a charged track,
and have energy greater than 50 MeV.
Pairs of photons with invariant mass within 3.5 $\sigma$
of the $\pi^0$ mass
(the $\gamma\gamma$ invariant mass rms resolution
$\sigma$ varies from 5 to 10 MeV/c$^2$, depending on $\pi^0$ energy)
are considered as $\pi^0$ candidates.
The energy of each $\pi^0$ is required to be greater than
$0.05 E_b$ or $0.10 E_b$, depending on tag.
Electrons and positrons are identified by requiring the
energy deposited in the calorimeter to be consistent
with the momentum of the charged track ($E/p > 0.8$),
and the energy deposited in the drift chamber (dE/dx)
to be no more than 2 $\sigma$ below that expected for electrons.
Muon candidates are required to penetrate $\ge$3 interaction lengths
of material in the muon system.

The $\pi^0$ candidates are associated with the charged track
nearest in angle, to form a $h^\pm\pi^0$ candidate.
The total visible energy in the event must be greater
than $0.3 E_{cm}$,
and the net transverse momentum in the event must exceed
$0.075 E_b$,
to suppress background from two photon processes.

Calorimeter showers that are not matched to charged tracks
or used in the $\rho$ candidates
can come from several sources, including
(a) decays of $\pi^0$'s from background
$\tau^+\tau^-$ events with higher $\pi^0$ multiplicity
(``feed-across'')
or from the hadronic continuum;
(b) photons from initial- or final-state radiation;
(c) secondary showers from hadronic interactions
in the calorimeter,
spatially separated from the parent, or
(d) showers from detector noise or the circulating beams,
but unrelated to the $e^+e^-$ interaction.
In order to suppress backgrounds from source (a),
we veto events which contain additional showers
having energy above 75 MeV, transverse energy profile
consistent with that for photons,
and $>$30 cm isolation from the entrance point
of any charged track into the calorimeter.
In the $3-\rho$ and $3-1$ topologies,
extra showers are permitted in the hemisphere
containing the 3-prong tag or 1-prong inclusive decay;
instead, the invariant mass of all observed particles
in these hemispheres is required to be less than the tau mass.
Modes containing an unobserved $K_L$ or $\omega\rightarrow \gamma\pi^0$
are considered to be $\tau^+\tau^-$ feed-across, {\it i.e.}, background
for which a correction will be applied.
The final branching fraction {\it does not include} such modes.

The loss of signal events due to sources (b) and (c)
is estimated using a detailed simulation
of the kinematics of signal topologies
from the KORALB \cite{ref:korb}
Monte Carlo event generator,
which generates radiative photons from $\tau^+\tau^-$
production and decay,
and of the
calorimeter response from a GEANT-based \cite{ref:geant}
detector simulation package.
The loss due to source (d)
is simulated by merging random-trigger events from the data
with Monte Carlo signal events.
The veto criteria have been designed to be
insensitive to the small differences between
Monte Carlo and data for hadronic showers,
which are more difficult to simulate than electromagnetic showers.


The number of events observed in each topology is given
in Table \ref{tab:events}.
Also listed
is the efficiency for reconstructing
the signal topologies (${\cal E}_S$),
estimated from Monte Carlo simulation,
and the fraction of observed events from the background ($f_B$).
The backgrounds from $\tau^+\tau^-$ feed-across
are estimated using Monte Carlo simulation
and world averages \cite{ref:PDG,ref:onepr,ref:npio} for the
branching fractions of the tau to feed-across modes
(dominantly from the decay $\tau^\pm \rightarrow h^\pm \pi^0\pi^0\nu_\tau$
\cite{ref:npio}
where one $\pi^0$ is not detected).


The efficiencies for reconstruction of signal topologies
and tau-pair backgrounds receive contributions from
the trigger, charged particle tracking,
photon detection and $\pi^0$ reconstruction,
and the effect of cuts on momenta, angles, and energies.
The trigger efficiency is greater than 99\%\ for all topologies
except for $e-\mu$ and $\mu-\rho$,
where the efficiency is as low as 95\%\ for some data-taking periods.
These efficiencies are measured using the data.
The charged-track-finding efficiency is also near 100\%\
for the track criteria used here;
this is verified by studying independent samples
of events with similar topologies.
The uncertainty in the track finding efficiency
is 0.5\%\ per track.
The lepton identification efficiencies and uncertainties cancel in the
$\ell-\rho$ method \cite{ref:leptonic}.
Fake lepton backgrounds are not included in $f_B$.

Reconstruction of $\pi^0$ decays is well simulated
by the GEANT-based Monte Carlo, as seen in Fig.~\ref{fig:pio}.
The peak position, width, low mass tail, and background level
are all well reproduced.
This is quantified by varying the cuts and methods used
to reconstruct photons and $\pi^0$'s, and determining
the change in the measured branching fraction.
Imperfect knowledge of
the detector material in front of the calorimeter
(and hence the photon conversion rate)
also contributes to uncertainty in the
absolute efficiency for $\pi^0$ reconstruction.

The distributions of all kinematical quantities in the signal events
are compared between data and Monte Carlo.
Examples of such comparisons are shown in Fig.~\ref{fig:kinem}.
There is good agreement in all distributions,
within the accepted region,
indicating that the acceptance is well modelled,
and that no significant background
(other than the $\tau^+\tau^-$ feed-across component included
in the Monte Carlo simulation) remains in the data.
Again, this is quantified by applying new cuts
or varying existing ones over a wide range in acceptance,
and observing the change in the measured branching fraction.
We have performed several other cross-checks of our $\pi^0$ reconstruction
and extra energy veto efficiency using our $\tau^+\tau^-$ data, all of
which yield consistent results \cite{ref:prd}.

The largest source of background in these topologies
is from $\tau^+\tau^-$ feed-across events in which the $h^\pm\pi^0$ system
is replaced by a system containing additional $\pi^0$'s which are
not observed.
The sensitivity of our result to
uncertainties in these branching fractions \cite{ref:npio}
is minimized by the tight veto on extra showers;
$f_B$ is small (see Table~\ref{tab:events}).
Backgrounds from tau decays containing
a $K_L$ or $\omega$ meson
can be estimated reliably from measured branching fractions \cite{ref:PDG}.
Other sources of backgrounds include 
tau decay modes
containing two spurious showers which fake a $\pi^0$;
our $\pi^0$ reconstruction algorithms reduce this background
to a negligible value.
The backgrounds from hadronic continuum and $B\bar B$ production
are very small except in the $3-\rho$ and $3-1$ topologies;
they are estimated from the data,
using events in which the invariant mass of all observed particles
in the 3-prong hemisphere is greater than the tau mass \cite{ref:3p2p0}.

The systematic errors associated with uncertainties in
the signal efficiency (${\cal E}_S$), the background fraction ($f_B$),
and the number of produced tau pairs ($N_{\tau\tau}$),
are tabulated for the three methods in Table~\ref{tab:syst}.
The largest sources of uncertainty are the absolute
$\pi^0$ reconstruction efficiency, the
inefficiency due to the
extra shower veto,
and the modelling of the inclusive 3-prong decay acceptance.

The values of ${\cal B}_{h\pi^0}$
determined using equation (1) are:
\begin{equation}
\begin{array}{rcl}
{\cal B}_{h\pi^0} & = & 0.2559 \pm 0.0019 \pm 0.0047  \quad (\ell-\rho); \\
{\cal B}_{h\pi^0} & = & 0.2567 \pm 0.0017 \pm 0.0045  \quad (\rho-\rho); \\
{\cal B}_{h\pi^0} & = & 0.2643 \pm 0.0029 \pm 0.0052  \quad    (3-\rho); \\
{\cal B}_{h\pi^0} & = & 0.2587 \pm 0.0012 \pm 0.0042  \quad (\hbox{combined}).
\end{array}
\end{equation}
The combined result is obtained
by weighting each measurement's statistical and
uncorrelated systematic errors.
Many of the largest sources of systematic errors
are common to all three methods,
reflected in the combined systematic error.
The three measurements have a $\chi^2$ of 2.4 for two degrees of freedom.


The $K^\pm\pi^0\nu_\tau$ component of our signal is obtained
from an independent measurement \cite{ref:kkk}
of the branching fraction
${\cal B}(\tau^\pm\rightarrow K^\pm \pi^0\nu_\tau) = 0.0051\pm0.0011$.
Thus, the $\pi^\pm\pi^0\nu_\tau$ mode
branching fraction is
${\cal B}_{\pi\pi^0} = 0.2536\pm0.0044$.
This compares well with a prediction \cite{ref:CVC}
derived from CVC and data on the cross section
$\sigma(e^+e^-\rightarrow\pi^+\pi^-)$,
${\cal B}_{\pi\pi^0} = 0.2458\pm0.0093\pm0.0027\pm0.0050$,
where the errors are from the $e^+e^-$ data, the tau lifetime,
and radiative corrections, respectively.

In conclusion, we have presented a high-precision measurement
of the branching fraction for the
decay $\tau^\pm\rightarrow h^\pm \pi^0 \nu_\tau$.
We have employed four tagging decays,
in three statistically independent combinations,
which are in good agreement with each another.
Common systematic errors limit the overall precision
to $\sim 1.7\%$ \cite{ref:prd}.

The measured value is larger and more precise
than previous measurements \cite{ref:PDG,ref:recent}.
This value,
in combination with resulting larger multi-$\pi^0$ ($h^\pm n\pi^0\nu_\tau$)
branching fractions \cite{ref:npio},
reduces the magnitude and significance of the one-prong problem.

We gratefully acknowledge the effort of the CESR staff in providing us with
excellent luminosity and running conditions.
This work was supported by the National Science Foundation,
the U.S. Dept. of Energy,
the Heisenberg Foundation,
the SSC Fellowship program of TNRLC,
and the A.P. Sloan Foundation.



%
\ifCLNS  
\input epsf.sty
\fi      

\begin{figure}[thb]
  \centering
\ifCLNS  
 \epsfysize6.54cm\leavevmode\epsfbox[70 200 446 500]{rf1n.ps}
\fi      
\medskip
\caption{The $M_{\gamma\gamma}$ distribution for the data (points)
and the Monte Carlo (histogram),
summed over all tags.}
\label{fig:pio}
\end{figure}

\begin{figure}[thb]
  \centering
\ifCLNS  
 \hbox{        \epsfysize5.50cm\leavevmode\epsfbox[76 200 460 500]{rf2an.ps}
   \hskip 5mm  \epsfysize5.50cm\leavevmode\epsfbox[76 200 460 500]{rf2bn.ps}}
               \epsfysize5.50cm\leavevmode\epsfbox[76 200 460 500]{rf2cn.ps}
\fi      
\medskip
\caption{
The distributions of a variety of kinematical variables
for the data (points) and the Monte Carlo (histogram),
summed over all tags.
(a) The scaled momentum of the charged particles, $p_\pi/E_b$
    (open circles and solid histogram), and
    the scaled energy of the $\pi^0$, $E_{\pi^0}/E_b$
    (solid circles and dotted histogram),
     scaled by a factor 3 for clarity);
(b) the scaled visible energy, $E_{vis}/E_{cm}$;
(c) the $\pi^\pm\pi^0$ invariant mass.
The accepted region is to the right of the vertical lines
in (a) and (b).}
\label{fig:kinem}
\end{figure}


\begin{table}[thb]
  \begin{center}
  \caption[]{For each event topology, we give
   the total number of events found,
   the background fraction ($f_B$),
   and the signal reconstruction efficiency (${\cal E}_S$).}
  \label{tab:events}
   \begin{tabular}{c|rrr}
Topology     &  Events    & $f_B$ (\%) &  ${\cal E}_S$ (\%) \\ \hline
$e-\rho$     &  13353     &  2.9      &   9.8             \\
$\mu-\rho$   &  11758     &  2.9      &   8.9             \\
$e-\mu$      &  16088     &  0.2      &  17.9             \\ \hline
$\rho-\rho$  &   6498     &  5.9      &   6.5             \\ \hline
$3-\rho$     &  12469     &  6.7      &  10.4             \\
$3-1$        &  85959     &  6.2      &  22.3             \\
   \end{tabular}
  \end{center}
\end{table}

\begin{table}[thb]
  \begin{center}
  \caption[]{Relative systematic errors (\%), for the three methods
             of measuring the branching fraction.}
  \label{tab:syst}
   \begin{tabular}{l|rrr}
Source of error            & $\ell-\rho$ & $\rho-\rho$ & $3-\rho$ \\ \hline
trigger efficiency         &  0.7    &  0.2    &  0.1   \\
tracking efficiency        &  0.5    &  0.5    &  0.2   \\
$\pi^0$ reconstruction     &  0.9    &  0.9    &  0.9   \\
acceptance                 &  0.5    &  0.5    &  1.1   \\
extra shower veto          &  0.9    &  0.9    &  0.9   \\
MC statistics              &  0.5    &  0.4    &  0.6   \\
$\tau^+\tau^-$ feed-across &  0.3    &  0.4    &  0.7   \\
non-$\tau$ backgrounds     &  0.2    &  0.4    &  0.4   \\
luminosity                 &  0.5    &  0.5    &  $-$   \\
$\sigma(\tau^+\tau^-)$     &  0.5    &  0.5    &  $-$   \\
${\cal B}_1$               &  $-$    &  $-$    &  0.4   \\ \hline
combined                   &  1.9    &  1.8    &  2.0    \\
   \end{tabular}
  \end{center}
\end{table}

\begin{references}

\bibitem{ref:PDG}
K.~Hikasa {\it et al.}, Phys.~Rev.~{\bf D45}, 1 (1992).

\bibitem{ref:recent}
Recent and preliminary results from
DELPHI, OPAL, ARGUS, and ALEPH were presented
by A.~Schwarz
at {\sl The XVI International Symposium
on Lepton Photon Interactions}, Cornell University,
August 1993,
MPI-PHE-93-24, Oct 1993.

\bibitem{ref:onepr}
See, for example, A.~Weinstein and R.~Stroynowski,
Ann.~Rev.~Nucl.~Part.~Sci.~{\bf 43}, 457 (1993).
We assume that the one-prong branching fraction ${\cal B}_1$
excludes modes containing $K_S\rightarrow\pi^+\pi^-$ decays;
the systematic error on ${\cal B}_1$ is increased
to account for this ambiguity.

\bibitem{ref:npio}
M.~Procario {\it et al.}, Phys.~Rev.~Lett.~{\bf 70}, 1207 (1993).

\bibitem{ref:CLEOd}
Y.~Kubota {\it et al.}, Nucl.~Instrum.~Methods {\bf A320}, 66 (1992).

\bibitem{ref:lumin}
G.~Crawford {\it et al.}, CLNS 94/1268 (1994), to be published in
Nucl.~Instrum.~Methods.

\bibitem{ref:korb}
KORALB (v.2.1) / TAUOLA (v.1.5):
S.~Jadach and Z.~Was, Comput.~Phys.~Commun.~{\bf 36}, 191 (1985)
and {\it ibid},~{\bf 64}, 267 (1991);
S.~Jadach, J.H.~K\"uhn, and Z.~Was,
Comput.~Phys.~Commun.~{\bf 64}, 275 (1991),
{\it ibid}, {\bf 70}, 69  (1992),
{\it ibid}, {\bf 76}, 361 (1993).

\bibitem{ref:geant}
R.~Brun {\it et al.}, GEANT 3.15, CERN DD/EE/84-1.

\bibitem{ref:leptonic}
These uncertainties are not negligible,
and precision measurements
of leptonic or three-prong inclusive branching fractions
cannot be obtained from the yields in Table \ref{tab:events}.

\bibitem{ref:prd}
Further details on the systematic errors, and
studies of the $\pi^\pm\pi^0$ invariant mass
and decay angle distributions, will be presented
in a future publication.

\bibitem{ref:3p2p0}
The method is described in D.~Bortoletto {\it et al.},
Phys.~Rev.~Lett.~{\bf 71}, 1791 (1993).

\bibitem{ref:kkk}
M.~Battle {\it et al.}, CLNS 94/1273 (1994),
submitted to Phys.~Rev.~Lett.

\bibitem{ref:CVC}
J.~K\"uhn and A.~Santamaria, Z.~Phys.~{\bf C48}, 443 (1990);
W.~Marciano, in
{\sl Proceedings of the Second Workshop on Tau Lepton Physics},
September 1992, ed.~K.K.~Gan, World Scientific (1993).

\end{references}
\end{document}